\documentclass[aps,prc,preprint,amsmath,showpacs,superscriptaddress,nofootinbib]{revtex4}
\usepackage{graphicx,epsfig,longtable}
\newcommand{\beqy}{\begin{eqnarray}}
\newcommand{\eeqy}{\end{eqnarray}}
\newcommand{\bmlet}{\begin{subequations}}
\newcommand{\emlet}{\end{subequations}}

\usepackage{epstopdf}

\begin{document}

\textwidth 16.2 cm
\oddsidemargin -.54 cm
\evensidemargin -.54 cm

\def\gsimeq{\,\,\raise0.14em\hbox{$>$}\kern-0.76em\lower0.28em\hbox  
{$\sim$}\,\,}  
\def\lsimeq{\,\,\raise0.14em\hbox{$<$}\kern-0.76em\lower0.28em\hbox  
{$\sim$}\,\,}  

\title{Masses of neutron stars and nuclei}
\author{N.~Chamel}
\affiliation{Institut d'Astronomie et d'Astrophysique, CP-226, Universit\'e
Libre de Bruxelles, 1050 Brussels, Belgium}
\author{A.~F.~Fantina}
\affiliation{Institut d'Astronomie et d'Astrophysique, CP-226,
Universit\'e Libre de Bruxelles, 1050 Brussels, Belgium}
\author{J.~M.~Pearson}
\affiliation{D\'ept. de Physique, Universit\'e de Montr\'eal, Montr\'eal
(Qu\'ebec), H3C 3J7 Canada}
\author{S.~Goriely}
\affiliation{Institut d'Astronomie et d'Astrophysique, CP-226,
Universit\'e Libre de Bruxelles, 1050 Brussels, Belgium}
\date{\today}

\begin{abstract}

We calculate the maximum mass of neutron stars for three different equations
of state (EOS) based on generalized Skyrme functionals that are simultaneously
fitted to essentially all the 2003 nuclear mass data (the rms deviation is 0.58
MeV in all three cases) and to one or other of three different equations of 
state of pure neutron matter, each determined by a different many-body 
calculation using realistic two- and three-body interactions but leading to 
significantly different degrees of stiffness at the high densities prevailing 
in neutron-star interiors. The observation of a neutron star with mass
1.97 $\pm$ 0.04 $\mathcal{M}_{\odot}$ eliminates the softest of
our models (BSk19), but does not discriminate between BSk20 and BSk21.
However, nuclear-mass measurements that have been made since our models were 
constructed strongly favor BSk21, our stiffest functional.
\end{abstract}


\maketitle

The nuclear energy-density functional theory aims at describing various nuclear
systems containing a large number of nucleons. It has been very successfully 
applied to the study of the structure and the dynamics of medium-mass and heavy
nuclei~\cite{bhr03}. This approach is also particularly well-suited to the
determination of the properties of dense nuclear matter in extreme 
astrophysical environments such as supernova cores and neutron-star 
interiors~\cite{sto07}. In this framework 
we recently constructed a family of three functionals, BSk19, BSk20 and 
BSk21~\cite{gcp10}, each one intended to provide a unified approach to the 
description not only of the different regions of neutron stars (including the 
possibility of superfluidity~\cite{cgpo10}) but also of supernova cores. These 
functionals have all been derived from Skyrme forces of the form
\beqy \label{1}
v_{ij} & = &
t_0(1+x_0 P_\sigma)\delta({\pmb{r}_{ij}})
+\frac{1}{2} t_1(1+x_1 P_\sigma)\frac{1}{\hbar^2}\left[p_{ij}^2\,
\delta({\pmb{r}_{ij}}) +\delta({\pmb{r}_{ij}})\, p_{ij}^2 \right]\nonumber\\
& + & t_2(1+x_2 P_\sigma)\frac{1}{\hbar^2}\pmb{p}_{ij}.\delta(\pmb{r}_{ij})\,
 \pmb{p}_{ij}
+\frac{1}{6}t_3(1+x_3 P_\sigma)\,n(\pmb{r})^\alpha\,\delta(\pmb{r}_{ij})
\nonumber\\
&  + & \frac{1}{2}\,t_4(1+x_4 P_\sigma)\frac{1}{\hbar^2} \left[p_{ij}^2\,
n({\pmb{r}})^\beta\,\delta({\pmb{r}}_{ij}) +
\delta({\pmb{r}}_{ij})\,n({\pmb{r}})^\beta\, p_{ij}^2 \right] \nonumber\\
& + &t_5(1+x_5 P_\sigma)\frac{1}{\hbar^2}{\pmb{p}}_{ij}.
n({\pmb{r}})^\gamma\,\delta({\pmb{r}}_{ij})\, {\pmb{p}}_{ij}
 + \frac{\rm i}{\hbar^2}W_0(\mbox{\boldmath$\sigma_i+\sigma_j$})\cdot
\pmb{p}_{ij}\times\delta(\pmb{r}_{ij})\,\pmb{p}_{ij}  \quad ,
\eeqy
where $\pmb{r}_{ij} = \pmb{r}_i - \pmb{r}_j$, $\pmb{r} = (\pmb{r}_i +
\pmb{r}_j)/2$, $\pmb{p}_{ij} = - {\rm i}\hbar(\pmb{\nabla}_i-\pmb{\nabla}_j)/2$
(this is the relative momentum), $P_\sigma$ is the two-body spin-exchange
operator, and $n(\pmb{r}) = n_n(\pmb{r}) + n_p(\pmb{r})$ is the total
local density, $n_n(\pmb{r})$ and $n_p(\pmb{r})$ being the neutron and
proton densities, respectively. These forces are generalizations of the 
conventional Skyrme form in that they contain terms in $t_4$ and $t_5$, which 
are density-dependent generalizations of the 
usual $t_1$ and $t_2$ terms, respectively~\cite{cgp09}.

\begin{figure}
\centerline{\epsfig{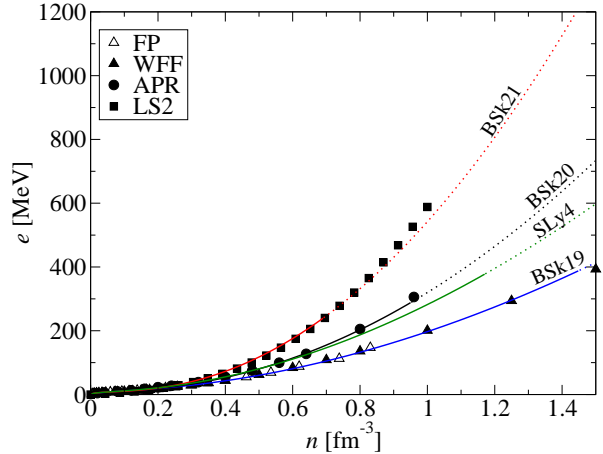}}
\caption{(Color online.) Zero-temperature EOSs for neutron matter (NeuM) with
functionals of this paper. Also shown are the realistic EOSs
FP~\cite{fp81}, WFF~\cite{wir88}, APR~\cite{apr98} and LS2~\cite{ls08}.
Dotted curves denote supraluminal EOS.}
\label{fig1}
\end{figure}

The parameters of this form of force were determined primarily by fitting
measured nuclear masses, which were calculated with the Hartree-Fock-Bogoliubov
(HFB) method. For this it was necessary to supplement the Skyrme forces with a
microscopic contact pairing force, phenomenological Wigner terms and correction
terms for the spurious collective energy. However, in fitting the mass data we
simultaneously constrained the Skyrme force to fit the zero-temperature
equation of state (EOS), i.e., the energy per nucleon $e$ as a function of the
density $n$, of homogeneous pure neutron matter (NeuM), as determined by 
many-body calculations with realistic two- and three-nucleon forces. Actually, 
several such calculations of the EOS of NeuM have been made, and while 
they all agree very closely at nuclear and subnuclear densities, at the much 
higher densities that can be encountered towards the center of neutron stars 
they differ greatly in the stiffness that they predict. It is in this way that
we arrived at the three
different effective forces of this paper: BSk19 was fitted to the softest
EOS of NeuM known to us, that of FP~\cite{fp81} and the one labeled 
``UV14 plus TNI" in Ref.~\cite{wir88} and which we refer to as WFF, BSk21 to 
the stiffest, the one labeled ``V18" in Ref.~\cite{ls08} and which we refer to 
as LS2, while BSk20 was fitted to an EOS of intermediate stiffness, the one 
labeled ``A18 + $\delta\,v$ + UIX$^*$" in Ref.~\cite{apr98} and which we refer 
to as APR (see Fig.~\ref{fig1}). Furthermore, the strength of the pairing force
at each point in the nucleus in
question was determined so as to exactly reproduce realistic $^1S_0$ pairing
gaps of homogeneous nuclear matter of the appropriate density and charge
asymmetry~\cite{cha10}. Finally, we imposed on these forces a number of 
supplementary realistic constraints, the most notable of which is the 
suppression of an unphysical transition to a spin-polarized configuration both 
at zero and finite temperatures, at densities found in neutron stars and 
supernova cores~\cite{cgp09,cg10,gcp10}. 

The introduction of the unconventional Skyrme terms allowed us to satisfy all 
these constraints and at the same time fit the 2149 measured masses of nuclei 
with $N$ and $Z \ge$ 8 given in the 2003 Atomic Mass Evaluation 
(AME)~\cite{audi03}
with an rms deviation as low as 0.58 MeV for all three models, i.e., for all
three options for the high-density behavior of NeuM. 
Fig.~\ref{fig1} also shows the EOS in NeuM of Skyrme force
SLy4~\cite{cha98}. This force, like our own, was intended for use in 
neutron stars. We see that its EOS in NeuM lies between those of BSk19 and
BSk20, i.e., between the realistic EOSs of FP-WFF and APR. It was also fitted
to a few nuclear masses, but the overall quality of the fit is rather poor,
an rms deviation of 5.1 MeV for all the measured even-even nuclei being 
quoted~\cite{dob04}. 

\begin{figure}
\centerline{\epsfig{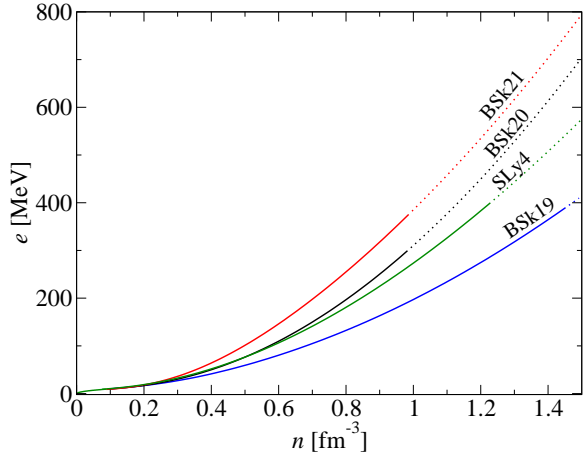}}
\caption{(Color online.) Zero-temperature EOSs for neutron-star matter (N*M) 
with functionals of this paper. Dotted curves denote supraluminal EOS.} 
\label{fig2}
\end{figure}
   
{\it Neutron-star structure.} With each of our functionals we have calculated 
the zero-temperature EOS in each of the three regions of neutron stars. In 
Ref.~\cite{gcp10} we calculated the EOS of neutron-star matter (N*M), the 
medium constituting the homogeneous core (assumed here to consist of just 
neutrons, protons, electrons and muons), which dominates the global properties
of neutron stars. Referring to Fig.~\ref{fig2},
we see that qualitatively, the EOS of N*M resembles that of NeuM for the same 
force, with BSk19 still having the softest EOS and BSk21 the stiffest.
Fig.~\ref{fig2} also shows the EOS of functional SLy4 in N*M. 

\begin{figure}
\centerline{\epsfig{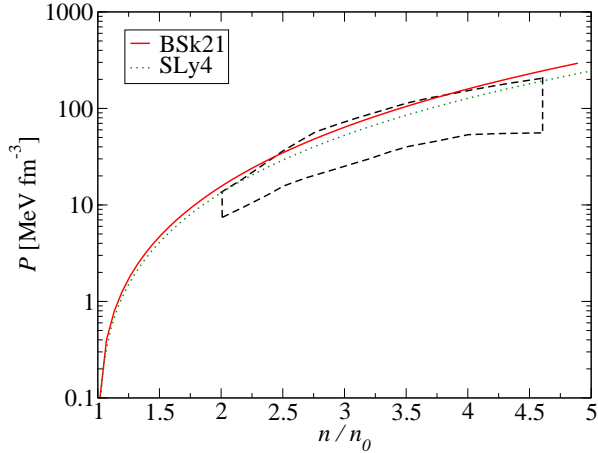}}
\caption{(Color online.) Calculated pressures in symmetric nuclear matter (SNM) for functionals 
of this paper (BSk19 and BSk20 yield almost undistinguishable results from BSk21 on this figure) ; the 
box summarizes the range of the measurements reported in Ref.~\cite{dan02}.}
\label{fig3}
\end{figure}

Aside from the obvious influence of the leptons, the
main reason for the differences between the EOS of NeuM and that of N*M lies
in the presence of protons, which interact both with each other and with 
neutrons. The precision mass fit assures that these interactions are well
represented, at least up to densities not much greater than $n_0$, the 
equilibrium density of symmetric nuclear matter (SNM). As for higher densities, 
the extent to which the EOS of N*M is not determined by that of NeuM depends on
the EOS of SNM, and as can be seen in Fig.~\ref{fig3}, 
our models are
consistent with measurements of the high-density pressure of SNM deduced from
heavy-ion collisions~\cite{dan02}, even though our functionals were not
directly fitted to the EOS of SNM. Thus, all in all, we believe that the 
procedure for determining our forces has an optimal mix of theory and 
experiment, at least as far as the EOS of N*M is concerned.

Our functionals are equally well adapted to the calculation of the EOS of the 
neutron-star crust. For the outer part of the crust, consisting of just 
electrons and bound nuclei, we used the HFB-19, HFB-20 and HFB-21 mass 
models~\cite{pgc11}, whenever experimental masses are not available.
These are based entirely on the BSk19, BSk20 and BSk21 
parameters, respectively, 
and were constructed to run from one drip line to the other~\cite{gcp10}.
Given not only the precision fit that these models give to the measured masses
but also the constraints to NeuM imposed on the underlying forces,
no other published mass model will make more reliable predictions for the 
experimentally inaccessible neutron-rich nuclei that dominate the outer crust.
(For the same reason these mass models are well adapted to the study of 
the r-process of nucleosynthesis \cite{sg11a,sg11b}.) For the inner crust, 
which consists of 
neutron--proton clusters embedded in a neutron vapor and electron gas, we have 
likewise calculated the EOS with each of the BSk19, BSk20 and BSk21 
functionals, using the ETFSI (extended Thomas-Fermi plus Strutinsky integral)
method~\cite{pcgd11}. Our confidence in the use of our functionals
in this region derives not only from the NeuM constraints to which
they have been subjected but also from the precision fit to masses, which means
that the presence of inhomogeneities and of protons is well represented.

For each of these three Skyrme functionals we thus have at our disposal the 
zero-temperature EOS for
all parts of neutron stars, and hence can solve the Tolman-Oppenheimer-Volkoff 
(TOV) equations describing the global structure of spherical non-rotating
neutron stars~\cite{tol39,ov39}. Solving these equations for different values 
of the baryonic central density, $n_{cen}$, gives us the total gravitational 
mass $\mathcal{M}$ as a function of the circumferential radius $R$. Numerical 
calculations have been carried out using the LORENE library~\cite{lorene}. 
The maximum possible value 
of $\mathcal{M}$, $\mathcal{M}_{max}$, along with the 
corresponding radius $R$ and central density $n_{cen}$, are given in the last 
three columns of Table~\ref{tab1} for each of our three functionals. The second
column of this table shows 
$n_{caus}$, the maximum value of the baryonic density in N*M for which 
causality holds, i.e., for which 
the velocity of sound is lower than the velocity of light, $c$ (see Section VB
of Ref.~\cite{gcp10} for details). It will be seen that for functionals BSk19 and
BSk20 the central density $n_{cen}$ for maximum mass is at the limit imposed by
causality; in fact if we had waived this constraint we would have found 
slightly higher (and unphysical) values of $\mathcal{M}_{max}$ and 
$n_{cen}$ (the maximum mass would have been essentially the same for BSk19, whereas it 
would have increased to $\mathcal{M}_{max}=2.17\mathcal{M}_{\odot}$ 
for BSk20). On the other hand, in the case of functional BSk21, which has the 
stiffest EOS, the requirement of causality is satisfied for all possible 
masses. Table~\ref{tab1} also shows (in parentheses), for each of BSk19, BSk20
and BSk21, the value of $\mathcal{M}_{max}$ that has already been published
for the corresponding realistic calculation, i.e., WFF, APR and LS2, respectively.    
The close agreement between corresponding values of $\mathcal{M}_{max}$ is a
measure of how well we have reproduced the realistic forces with our
Skyrme functionals. 

\begin{table*}
\centering
\caption{For each functional are shown i) limiting baryonic density for 
causality, ii) maximum neutron-star mass
(the quantities in parentheses refer to the corresponding realistic forces)
iii) corresponding radius and iv) corresponding central baryonic density.
(Neutron-star results for SLy4 taken from Ref.~\cite{dh01}.)}
\label{tab1}
\vspace{.5cm}
\begin{tabular}{|c|cccc|}
\hline
Force &  $n_{caus}$ (fm$^{-3}$) &   $\mathcal{M}_{max}/\mathcal{M}_{\odot}$ & $R$ (km)& $n_{cen}$ (fm$^{-3}$)\\
\hline
BSk19&  1.45 &  1.86 (1.84 \cite{wir88})&  9.13 & 1.45  \\
BSk20&  0.98 &  2.15 (2.20 \cite{apr98})&  10.6 & 0.98   \\
BSk21&  0.99 &  2.28 (2.3 \cite{ls08})& 11.0 & 0.98   \\
SLy4 & 1.23  & 2.05 & 9.99 & 1.21 \\ 
\hline
\end{tabular}
\end{table*}

Comparing the values of $\mathcal{M}_{max}$ shown in Table~\ref{tab1} with the 
recently measured value of 1.97 $\pm$ 0.04 $\mathcal{M}_{\odot}$ for the mass 
of pulsar PSR J1614$-$2230~\cite{dem10} shows that the EOS of BSk19 for N*M and 
NeuM is definitely too soft. Even taking account of the fact that this pulsar 
rotates with a frequency of 317 Hertz, we have found using the LORENE library~\cite{lorene} 
(considering stationary configurations for rigidly rotating neutron stars, 
see, e.g. Ref.~\cite{gg10}) that $\mathcal{M}_{max}$ is raised only by 
0.005~$\mathcal{M}_{\odot}$, which is insufficient 
to change this conclusion. On the other hand, functionals BSk20 and BSk21, along 
with SLy4, all provide EOSs that are sufficiently stiff to support PSR 
J1614$-$2230. 

\begin{figure}
\centerline{\epsfig{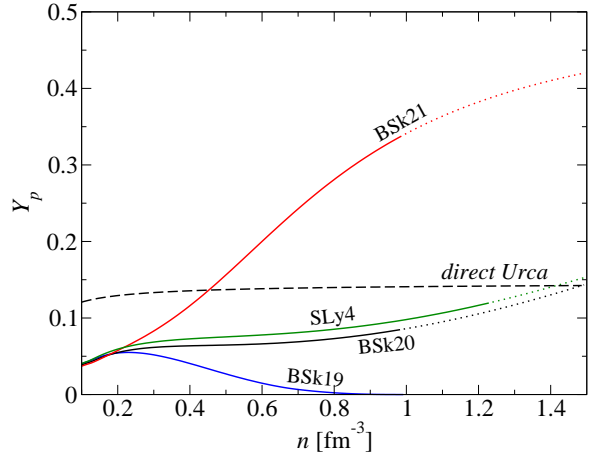}}
\caption{(Color online.) Proton fraction $Y_p$ in neutron-star matter for
functionals of this paper. The dashed line indicates the threshold value of 
$Y_p$ for the direct Urca process to be allowed. Dotted curves denote 
supraluminal EOS.} 
\label{fig4}
\end{figure}

Referring to Fig.~\ref{fig4}, we see that BSk21 is the only one of our 
functionals for which the proton fraction can exceed the critical threshold
for a  direct Urca process to occur, and then only for baryon density $n>0.45$ 
fm$^{-3}$, i.e., only in neutron stars whose gravitational mass 
$\mathcal{M}>1.59\mathcal{M}_\odot$. As a consequence, our three functionals 
are compatible with the constraint of Ref.~\cite{kla06} that no direct Urca 
process should occur in neutron stars with masses in the range 1 - 1.5~$\mathcal{M}_\odot$. 
However, we now see that recently accumulated nuclear-mass data strongly favor BSk21.

{\it Nuclear Masses.} Since we fitted our forces to the data of the 2003 AME 
many new mass data have been accumulated. In particular, a new unpublished AME 
became available in 2011~\cite{ame2011}. This contains 154 new measurements in 
the interval $N$ and $Z \ge$ 8, while 9 masses that were previously listed as 
``measured" have now been demoted to an ``estimated" status; in accordance with
our previous practice we do not consider such nuclei.

The first three lines of Table~\ref{tab2} show the rms and mean deviations of 
our HFB-19, HFB-20 and HFB-21 mass models with respect to all measured masses
in the interval $N$ and $Z \ge$ 8 in the 2003 AME, while the next three lines
show the corresponding results for the subset consisting of nuclei with a 
neutron separation energy of less than 5.0 MeV. A slight favoring of HFB-21 
over the other two models is already apparent in this neutron-rich domain, the 
region of interest in neutron-star applications. This trend is strengthened 
somewhat in the next six lines, where we show the corresponding results for the
2011 AME. Accordingly, in the next three lines (13 to 15) we look at the 
deviations with respect to just the 154 new data points: the discrimination in 
favor of HFB-21 is now quite marked, although both neutron-rich and proton-rich
nuclei are included in this set. Of these 154 new nuclei 73 lie on the 
neutron-rich side of the stability line, and we show the corresponding 
deviations in lines 16 to 18 of Table~\ref{tab2}. There is now an even 
stronger selection of HFB-21.

Not included in either AME are the 41 unpublished mass measurements of
highly neutron-rich nuclei by Matos~\cite{mat04}. The 2011 AME lists 24 of 
these nuclei, and although the Matos data were not used the agreement is 
very good. We consider the subset of the remaining 17 Matos data in the last 
three lines of Table~\ref{tab2}, where we
see further confirmation of a strong preference for HFB-21. 

\begin{table*}
\centering
\caption{Rms and mean deviations between mass models and different data sets
(see text).      }   
\label{tab2}
\vspace{.5cm}
\begin{tabular}{|ccc|cc|}
\hline
Data set& Number of nuclei &Mass model & rms (MeV)&${\bar{\epsilon}}$ (MeV)\\
\hline
2003 AME & 2149&HFB-19& 0.583 & -0.038\\ 
" & " & HFB-20&0.582 & 0.021\\
" & " & HFB-21&0.577 & -0.054\\
\hline
2003 AME n-rich & 185&HFB-19& 0.803 & 0.243\\
" & " & HFB-20&0.790 & 0.217\\
" & " & HFB-21&0.762 & -0.086\\
\hline
2011 AME & 2294&HFB-19& 0.595 & -0.010\\
" & " & HFB-20&0.593 & 0.051\\
" & " & HFB-21&0.574 & -0.031\\
\hline
2011 AME n-rich & 224&HFB-19& 0.799 & 0.261\\
" & " & HFB-20&0.788 & 0.255\\
" & " & HFB-21&0.730 & -0.060\\
\hline
2011 new& 154&HFB-19& 0.824 & 0.310\\
" & " & HFB-20&0.803 & 0.375\\
" & " & HFB-21&0.681 & 0.185\\
\hline
2011 new n-rich  & 73&HFB-19& 0.948 & 0.391\\
" & " & HFB-20&0.893 & 0.408\\
" & " & HFB-21&0.735 & 0.063\\
\hline
2007 Matos& 17&HFB-19& 0.912& 0.757\\
" & " & HFB-20&0.878 & 0.713\\
" & " & HFB-21&0.536 & 0.216\\
\hline
\end{tabular}
\end{table*}

{\it Conclusions.} Our three nuclear mass models HFB-19, HFB-20 and HFB-21, 
which are based on the functionals BSk19, BSk20 and BSk21, respectively, give 
equally good fits to the nuclear mass data appearing in the 2003 
AME~\cite{audi03}. However, since 2003, and since our mass models were fitted, 
mass measurements have begun to be made sufficiently far from the stability 
line to discriminate between our models. Specifically, model BSk21 is strongly
favored by the new data. 

To complement this conclusion we have solved the general-relativistic equations
of rotating and non-rotating stars for our three 
functionals and find that BSk19 is too soft at high densities to support 
neutron stars as heavy as the heaviest yet observed, while BSk20 and BSk21 
will. This corroboration between measurements of nuclear masses and
measurements of neutron-star masses is gratifying, but it should not be
concluded that nuclear-mass measurements are now tying down the high-density
EOS of N*M, since with a more elaborate Skyrme parametrization (e.g., by taking
several forms of each of the $t_3, t_4$ and $t_5$ terms) it might be possible to 
have an equally good fit to all the mass data while
making different extrapolations to high-density N*M. Nevertheless, at the very 
least we recognize that in BSk21 we have a functional that performs well over
a wide range of densities, giving a precision fit to nuclear masses, and being
compatible with what is now known about neutron-star masses.

We would like to thank M. Matos for sending us his thesis results, and
G. Audi, P. Haensel and D. Lunney for valuable communications.
The financial support of the FNRS (Belgium), the NSERC (Canada) and CompStar 
(a Research Networking Programme of the European Science Foundation) is
gratefully acknowledged.

\end{document}